# A multi-stage, first-order phase transition in LaFe$_{11.8}$Si$_{1.2}$: interplay between the structural, magnetic and electronic degrees of freedom

K.P. Skokov$^{1,*}$, A.Y. Karpenkov$^1$, D.Y. Karpenkov$^1$, I.A. Radulov$^1$, D. Günzing$^2$, B. Eggert$^2$, A. Rogalev$^3$, F. Wilhelm$^3$, J. Liu$^{4,5}$, Y. Shao$^4$, K. Ollefs$^2$, M. E. Gruner$^2$, H. Wende$^2$, and O. Gutfleisch$^1$

$^1$Institute of Materials Science, Technical University of Darmstadt, Darmstadt 64287, Germany
$^2$Faculty of Physics, University of Duisburg-Essen, D-47048 Duisburg, Germany
$^3$European Synchrotron Radiation Facility, Grenoble F-38043, France
$^4$Key Laboratory of Magnetic Materials and Devices, Ningbo Institute of Material Technology and Engineering, Ningbo 315201, China
$^5$School of Materials Science and Engineering, Shanghai University, Shanghai 200444, China

Alloys with a first-order magnetic transition are central to solid-state refrigeration technology, sensors and actuators, or spintronic devices. The discontinuous nature of the transition in these materials is a consequence of the coupling between the magnetic, electronic and structural subsystems, and such transition can, in principle, cross several metastable states, where at one point the transition takes place within the magnetic subsystem, while at another the changes occur in the structural or electronic sub-systems. To address this issue, we conducted simultaneous measurements of the *macroscopic* properties – magnetization, temperature change of the sample, longitudinal and transversal magnetostrictions – to reveal the rich details of the magneto-structural, first-order transition occurring in the prototypical alloy LaFe$_{11.8}$Si$_{1.2}$. We found that the transition does not complete in one but in two distinct stages. The presence of the intermediate state changes the potential-energy landscape, which then impacts strongly on the width of the hysteresis associated with the first-order transition. We complement these findings with experiments on the *atomistic* scale, i.e., x-ray absorption spectroscopy (XAS), x-ray magnetic circular dichroism (XMCD) and Mössbauer spectroscopy, and then combine them with first-principles calculations to reveal the full complexity and two-stage nature of the transition. This new approach can be successfully extended to a large class of advanced magnetic materials that exhibit analogous transformations.

## I. INTRODUCTION.

Magnetic materials that respond to multiple stimuli depending on their structural, magnetic, and electronic degrees of freedom are at the forefront of research in material physics [1–4]. They form a new family of materials for technological areas like heat-assisted magnetic recording, thermal energy storage, novel spintronic devices and magnetic refrigeration [5–7]. Large responses can be expected for materials with a first-order phase transition, where the application of different thermodynamic fields is accompanied by large discontinuities in their conjugate variables [8,9]. With such materials the changes caused by one external field in one subsystem of the solid (e.g., the change of magnetization induced by an applied magnetic field) immediately gives rise to transformations in other subsystems (e.g., an expansion of the crystal lattice, a change in the electrical resistivity, an increase in the temperature of the sample) [10–12]. Therefore, it is important to understand which subsystem triggers the phase transition, the interplay of the different subsystems of the solid, and how these impact on the resulting caloric effects [13–15]. Moreover, to minimize hysteresis losses, an understanding of the interplay of the involved degrees of freedom on the macroscopic, atomistic and electronic length scales is a prerequisite.

Despite the complexity of such phase transitions, the thermodynamic models that are commonly employed to describe this phenomenon usually assume that the transformation occurs in a single step. Thus, for an isostructural transition, the Kittel [16], Bean-Rodbell [17] and generalized Landau-Ginsburg models [18,19] presume a linear dependence of the exchange energy on the lattice parameters, which inevitably leads to the single-stage, field-induced transformation of a paramagnetic (or antiferromagnetic) phase to a ferromagnetic one. Meanwhile, due to the strong interplay between the structural, magnetic, and electronic degrees of freedom, such a first-order transition can, in principle, cross several metastable states, where at one point the transition takes place within the magnetic subsystem, while at another the changes occur in the structural or electronic sub-systems [20–22].

To quantify the field-, stress- and temperature-driven effects, researchers measure the changes taking place in a single subsystem of the solid. Conventional magnetometry [23,24], the adiabatic temperature change $\Delta T_{ad}$ [25,26], calorimetry in a magnetic field [27,28], conventional dilatometry or X-ray/neutron diffraction under an applied magnetic field [29–32] are some of the standard experimental setups in various laboratories. This leads to the situation where the data in the literature were not always acquired under comparable conditions (for instance, samples with different demagnetization factors, various field-sweeping rates, different temperatures in the cryostat and on the sample, etc). Moreover, some measurements are performed on powdered samples, whereas others are carried out on single crystals, polycrystalline bulk samples or thin films. This hampers the subsequent use of the data in theoretical analyses and quantitative modelling, making it extremely difficult to assemble the various pieces of the puzzle and provide a complete picture of the magneto-structural, first-order phase transition, and probably the reason why the mechanisms of magneto-structural coupling remain unresolved. Thus, for the rational design of these materials, it is vitally important to know in detail, how different subsystems of the solid interplay during the transition, which system triggers the phase transformation, and how this mutual entanglement interaction can be responsible for the resulting thermal effects.

Therefore, it is important to change the existing paradigm and begin characterizing magnetic phase-change materials with a combination of experimental techniques, which means at least: (1) field/temperature dependencies of magnetization; (2) changes occurring in the crystal lattice; (3) thermal response of the material. However, the key point is that all these data must be collected simultaneously from a single sample and under identical experimental conditions. Only then can we understand the interplay of the degrees of freedom and tailor the material for a particular application.

One of the most promising classes of magnetocaloric material is the La(Fe,Si)$_{13}$-type compounds [33–35]. In the compositional range $1.2<x<1.6$ the LaFe$_{13-x}$Si$_x$ phase undergoes a first-order, itinerant electron meta-magnetic (IEM) transition accompanied by a large change in volume [36–38]. Being much more abundant and thus much cheaper than the benchmark magnetocaloric material Gd, La(Fe,Si)$_{13}$-type alloys are very attractive for commercial use as a magnetic refrigerant [39,40]. Their narrow magnetic and temperature hystereses make it possible to utilize the entropic advantages of a first-order transition without a significant reduction of the magnetocaloric effect (MCE) due to hysteretic losses [41,42] and so construct effective magnetic cooling devices.

In this paper we demonstrate how simultaneous measurements of the *macroscopic* properties, such as magnetization, temperature change, longitudinal and transversal magnetostriction, reveal rich details of the magneto-structural, first-order transition occurring in LaFe$_{11.8}$Si$_{1.2}$. We complement this with experiments on an *atomistic* scale: x-ray absorption spectroscopy (XAS), x-ray magnetic circular dichroism (XMCD) and Mössbauer spectroscopy, and combine them with first-principles calculations to reveal the two-stage nature of the transition as a direct consequence of IEM.

## II. METHODS.
### A. Materials

LaFe$_{11.8}$Si$_{1.2}$ ingots of 20 g were produced by induction melting from pure elements. After melting, to ensure homogeneity, the samples were encapsulated in quartz tubes under an Ar atmosphere and then annealed at 1050°C for 3 h. The homogenized ingots were then segmented into pieces for subsequent suction casting. Such casting solves two main problems. The first one is the net shaping of plates of 2×2×1 mm$^3$ required for the simultaneous measurement of magnetization, magnetostriction and thermal response. The second advantage is the reduction of annealing time from one week (typical annealing time for bulk arc-melted or induction-melted samples) to 12 h for suction-cast plates because of the more rapid diffusion resulting from the smaller grain size after quenching. The suction casting of 1 g samples involved a rectangular copper mould of 10×4×0.5 mm$^3$.

The suction-cast samples consisted predominantly of $\alpha$-Fe and La-rich phases, and therefore, an additional annealing step is required to obtain the magnetocaloric 1:13 phase, formed by a peritectic reaction. The suction-cast samples were wrapped in Mo foil, sealed in quartz tubes, annealed in a resistance tube furnace at 1100°C for 12 h, and subsequently quenched in water.

### B. Characterization.

Microstructural characterization and average bulk compositional quantification of the samples were performed using a Philips XL30 FEG scanning electron microscope equipped with an energy-dispersive x-ray system (SEM/EDX). The x-ray diffraction (XRD) patterns of the pulverized samples were collected at room temperature using a Stoe Stadi P instrument in transmission mode with a Mo $K\alpha_1$ source of radiation. Rietveld refinement of the XRD pattern was accomplished using the FullProf program. Analysis showed that the annealed LaFe$_{11.8}$Si$_{1.2}$ annealed plates contain 95% of the desired 1:13 phase and approximately 4% of $\alpha$-Fe and 1% of nonmagnetic La-Fe-Si phases.

### C. Adiabatic simultaneous measurements.

Direct measurements of the adiabatic temperature change $\Delta T_{ad}$ and the longitudinal and transversal adiabatic magnetostrictions were performed with an in-house experimental setup (Fig 1). This setup was built on our previous adiabatic temperature measurement device detailed in [25,26,35]. The magnetic field was produced by permanent magnets (Halbach type, AMT&C) arranged in two concentric cylinders rotating in opposite directions. The magnetic field could be changed in the centre of the bore from 1.93 T to -1.93 T at a rate of about 1 T/s, fast enough to ignore the heat losses from the sample to the environment during the measurement. The temperature changes of the sample were measured using a copper-constantan thermocouple (T-type) with an accuracy better than ±0.01 K. Due to the high vacuum in the chamber and thermal insulation (cryogel), the temperature across the sample was uniform with an accuracy of ±0.02 K. Four strain gauges (The Vishay Micro-Measurements strain gauges SK-06-031CF-350) were used for measurements of magnetostriction: two strain gauges for measurements of the longitudinal and transversal magnetostrictions were glued to the sample, and two more were glued on the sapphire plate with known thermal expansion coefficient for subtracting the effects of the field and the temperature on the strain gauge's resistivity. For more accuracy, the strain gaugeы уку connected to a Wheatstone bridges. The voltage-fed Wheatstone bridges were then compensated before each measurement. Despite the fact that modern strain gauges allow measurements of magnetostriction and thermal expansion with an accuracy of several ppm, we carried out calibration measurements using single crystals of nickel, gadolinium and Terfenol-D. These calibration samples were also measured on a standard dilatometer and both measurements showed agreement within 0.1%.

### D. Near isothermal simultaneous measurements.

For the simultaneous measurements of magnetization, magnetostriction and temperature change of the $LaFe_{11.8}Si_{1.2}$, we used a purpose-built experimental setup (Fig 2). This scientific instrument greatly enhances the measurement capacity of the commercial VSM option for the QD PPMS. Two strain gauges were glued on the sample surface by using M-Bond 610 Adhesive for measurements of the longitudinal and transversal magnetostrictions. To ensure accurate measurement, the strain gauge grids covered the entire surface of the sample. Two additional strain gauges were glued on a sapphire plate and placed near the sample to correct the changes in the strain gauges' resistivities induced by the external magnetic field and temperature. For the sample temperature, resistive sensors Cernox CX-1050-BC were welded to the sample with indium.

Knowledge of the sample temperature $T_{sample}$ gives us a unique opportunity to compare the moments when the main heat transfer takes place with the changes in the sample's dimensions and magnetization. Another undeniable advantage of our measurement technique is that the temperature of the sample $T_{sample}$ is measured directly, whereas in all commercially available devices, intended for measurement of magnetization, thermal expansion or resistance, only the temperature in the cryostat is taken into account, which, in principle, can differ from the sample temperature by 1-2 K. Since the temperature in the cryostat is taken as the temperature of the sample in conventional PPMS measurements, the method developed by us also makes it possible to more accurately determine the width of the temperature hysteresis and the transition temperature.

### E. XAS/XMCD measurements

The XAS/XMCD measurements at the Fe K-edge were performed at the ID12 beamline of the European Synchrotron Radiation Facility (ESRF). The APPLE-II-type undulator was used as a source of circularly polarized photons in the required energy range. The experimental station is equipped with a 17-T superconducting magnet and the sample temperature is controlled by a constant-flow cryostat operating between 2.5 and 325 K. The XAS measurements were carried out on a piece of $LaFe_{11.8}Si_{1.2}$ using total-fluorescence-yield detection in a backscattering geometry. All the spectra were corrected for fluorescence re-absorption effects. The XMCD spectra were obtained as the difference between the XAS spectra recorded with opposite helicities of the incoming X-ray beam and the magnetic field applied collinearly with the photon propagation direction. The isotropic XAS spectra correspond to the average of two spectra with opposite helicities. To ensure an XMCD signal free of artifacts, the measurements are repeated for the opposite direction of the magnetic field. The resulting XMCD spectrum corresponds to the average of only three XAS differences for each magnetic field direction. To improve the signal-to-noise ratio of the element selective magnetization curve we plotted the integrals of the absolute values of the XMCD signals, instead of a single point of the XMCD spectra as a function of applied field.

### F. Mössbauer spectroscopy

$^{57}$Fe Mössbauer spectroscopy measurements on the of $LaFe_{11.8}Si_{1.2}$ powder sample were conducted in the transmission geometry with a zero external magnetic field. A $^{57}$Co source (Rh matrix) and a constant-acceleration spectrometer were used, combined with conventional electronics. The low-temperature measurements were performed in a liquid-helium-bath cryostat. All samples had natural isotopic abundance (~2% 57Fe). The Mössbauer spectra were least-squares fitted using the computer program package 'Pi' by von Horsten.

### G. DFT simulations

The DFT calculations were carried out with the Vienna Ab-initio Simulation Package (VASP) [43,44]. For the exchange-correlation functional we employed the generalized gradient approximation (GGA) in the formulation by Perdew and Wang PW91 [45,46] in connection with the spin interpolation formula of Vosko, Wilk, and Nusair [47]. We used the scalar-relativistic approximation with a collinear spin setup, while the metastable disordered magnetic states were realized by employing the fixed-spin-moment (FSM) method [48] using the scheme described in [8,13].

We used projector-augmented-wave (PAW) [49] potentials with a valence electron configuration of $5s^25p^65d^16s^2$ for La, $3d^74s^1$ for Fe, and $3s^23p^2$ for Si. The plane wave cutoff for the electronic structure calculations was 335 eV. For the structural optimizations (cell parameter and atomic positions) in the 112-atom cell we used a k-mesh of 2×2×2 Monkhorst–Pack grid a finite temperature smearing according to Methfessel and Paxton [50] with a broadening of 0.1 eV. The self-consistency cycle was stopped when the difference in energy between two consecutive cycles fell below $10^{-7}$ eV and the optimized structures had residual forces of less than $5 \times 10^{-3}$ eV Å$^{-1}$.

### III. RESULTS AND DISCUSSION.
### A. Adiabatic measurements

Adiabatic magnetization occurs without the transfer of thermal energy to the environment, keeping the total entropy of the system unchanged. At the same time, the redistribution of the entropy between the lattice, magnetic and electronic subsystems of the solid leads to a change in the sample temperature, and near the phase transition such changes achieve their maxima. It is known that the field-induced transition from paramagnetic to ferromagnetic state in $LaFe_{11.8}Si_{1.2}$ is accompanied by a large magnetocaloric effect along with a significant increase in the volume (up to 1%). Thus, in the context of this work, it is very important to know whether a change in temperature will occur synchronously with a change in volume (which is implied in many theoretical approaches) or whether this is a more complex and non-linear process.

Simultaneous measurements of the adiabatic temperature change $\Delta T_{ad}$ and both the longitudinal $\Delta l_{ad,\parallel}(H)/l_{\parallel,0}$ and transversal $\Delta l_{ad,\perp}(H)/l_{\perp,0}$ adiabatic magnetostrictions were performed in our specifically designed setup. Since the $LaFe_{11.8}Si_{1.2}$ studied in this work has a cubic crystal structure, the magnetostriction is practically identical in all orientations of the sample. Using the longitudinal and transversal magnetostrictions, the volume-strain $\omega$ can be expressed as $\omega = 2\Delta l_{ad,\perp}(H)/l_{\perp,0} .+_{,\parallel}(H)/l_{\parallel,0}$.

Fig 3 (a,b) shows the field dependencies of the adiabatic temperature change $\Delta T_{ad}(H)$ together with the adiabatic volume change $\omega(H)$ for a $LaFe_{11.8}Si_{1.2}$ alloy in the vicinity of the first-order transition. Both $\Delta T_{ad}(H)$ and $\omega(H)$ were measured simultaneously under a magnetic field change $\Delta \mu_0 H = 1.9$ T, and the temperature in the cryostat was maintained at 187 K (Fig 3(a)) and 181 K (Fig. 3(b)). When the magnetic field is applied at 187 K (right-hand side of Fig. 3(a), $H>0$), in fields below 1 T the changes in sample temperature and volume are insignificant, since the sample is in the paramagnetic state. In fields above 1 T, a magneto-structural transition from the paramagnetic to the ferromagnetic state occurs, accompanied by heating of the sample with a slight reduction in its volume. When the field is reduced back to zero, both temperature and sample volume return to their initial values. Thus, at 187 K, the magnetization reversal losses, which are an important characteristic of first-order transitions, are practically absent, and the magnetiovolume effect has an anomalous sign.

When the magnetic field is applied at 181 K (Fig. 3(b)), $\Delta T_{ad}(H)$ gradually increases, whereas $\omega(H)$ remains almost unchanged until the magnetic field reaches $\mu_0 H=1.1$ T (first stage of the transition). In a higher field (second stage of the transition, upwards field leg) the $LaFe_{11.8}Si_{1.2}$ sample begins to expand abruptly, together with a continuous increase in $\Delta T_{ad}(H)$. During the backward transition that occurs when reducing the magnetic field, the cooling of the sample and the lattice contraction begin at the same time, but $\omega(H)$ decreases much faster than the temperature, and in magnetic fields below 0.5 T the temperature change occurs without changes to the dimensions of the sample. In the second cycle (left-hand side of Fig 3(b)), where the magnetic field is applied in the opposite direction, the two stages of the transition can be seen with equal clarity. It is important to note that in the case of a two-stage transition, after the first cycle of magnetization and demagnetization, the sample temperature does not return to its initial value (irreversible heating). Since the second stage is associated with transformations in the structural subsystem of the sample, it can be said that the structural transition is the cause of irreversible losses.

In order to show the overall picture, the temperature dependencies of $\Delta T_{ad}(H)$ and $\omega(H)$ measured simultaneously under a magnetic field change $\Delta\mu_0 H=1.9$ T during cooling and heating depicted in Fig 3 (c). In the temperature range 185–190 K (yellow shadowed area in Fig 3(c)) the magnetocaloric heating of the sample occurs without any lattice expansion. Instead, the sample exhibits a small reduction in its volume when the magnetic field is applied. In contrast, between 180K and 185K (green shadowed area in Fig 3(c)) the adiabatic temperature change of the sample is accompanied by the field-induced lattice expansion.

To summarize these findings, in Fig 3(d) we plotted thermal expansions $\omega(T)$ measured in zero field (dark-cyan lines) and in a field of 1.9 T (blue lines). The vertical dotted arrows show the isothermal field dependencies of $\omega(T)$ measured in fields up to 1.9 T. Under adiabatic conditions, the magnetic field is responsible for changes to both the dimensions of the sample and its temperature. Hence, in Fig 3(d) the dark-blue curves represent nonlinear adiabatic tracks $\omega_{ad}(H)$ started at 181 K and

187 K, showing the path on the $\omega$-$T$ diagram, where the system goes from a state with $\mu_0H=0$ and $T=T_0$ to a state with $\mu_0H=1.9$ T and $T=T_0+\Delta T_{ad}$, followed by its return to the zero-field state. At $T_0 = 187$ K, the first stage of the transition is characterized by heating of the sample without a significant change in its dimensions (lattice contraction is a consequence of negative thermal expansion coefficient). However, for $T_0=187$ K one can see the two-stage transition, and during the second stage, the heating/cooling and the expansion/contraction occur congruently, which leads to a change in the slope of the $\omega_{ad}(H)$ tracks.

### B. Near-isothermal measurements.

Thus, we see that at the first order transition the change in the temperature of the LaFe$_{11.8}$Si$_{1.2}$ and the expansion/contraction of the lattice (isostructural transformation) do not mirror each other, which leads to a two-step transition. However, to comprehend the interaction between the magnetic and structural subsystems, it is necessary to add magnetization to the consideration. Magnetization, the most important parameter, is difficult to measure under adiabatic conditions, and practically all commercial magnetometers measure $m(H)_T$ in near-isothermal conditions (e.g. the temperature in the cryostat is maintained constant, but if the material has a sufficiently large magnetocaloric effect, then during the transition induced by the magnetic field, the sample can heat up cool / down by 0.5-1K, which in principle is not perfect isothermal conditions). For such near-isothermal simultaneous measurements of magnetization, magnetostriction and temperature change of LaFe$_{11.8}$Si$_{1.2}$, we use our purpose-built experimental setup, based on a commercial PPMS system (VSM option).

Fig 4(a) shows the temperature dependencies of the specific magnetization $m(T)$. We used this data (measured up to 400 K) to plot the temperature dependence of the high-field inverse magnetic susceptibility $\frac{\mu_0}{\chi}$. Using $m(T)$, and $\frac{\mu_0}{\chi}(T)$ we estimated the magnetic moments per Fe atom in the ferromagnetic state $p_s=1.96$ $\mu_B$ and in the paramagnetic state $p_c=1.12$ $\mu_B$. It should be noted that such a low $p_c/p_s$ ratio is not consistent with the empirical rule of Rhodes and Wohlfarth [51], indicating fundamental changes in the electronic subsystem during the first-order phase transition.

The volume change $\omega(T)$ simultaneously measured in magnetic fields of 0.02, 2, 6, 10 and 14 T are shown in Fig 4(b). We can see from the inset in Fig 4(b) that in 0.02 T the changes in the magnetization and volume of the sample occur in two clearly distinguishable stages. For example, under cooling, the magnetization firstly increases together with a small contraction of the sample's volume, and in the second stage, an abrupt lattice expansion accompanies the subsequent increase in magnetization.

The specific heat $c_P(T)$ shown in Fig. 4(c) was measured for the same sample and in the same magnetic fields, confirming that during the phase transition an absorption/rejection of the thermal energy proceeds in two stages (see inset). With an increase in the magnetic field the transition temperature of LaFe$_{11.8}$Si$_{1.2}$ shifts upwards at a rate of 4.5 K/T, and the transition changes from first-order (0–8 T, 182–215K) to a continuous change (second-order), separated by a critical point [52].

A local-moment theory of volume magnetostriction was developed by Callen and Callen [53]. In the mean-field picture, the spontaneous volume magnetostriction is then proportional to the square of the magnetization: $\omega = kC_{mv}m^2$ ($k$: compressibility, $C_{mv}$: magneto-volume coupling constant). A similar relation was proposed by Shimizu for the spontaneous magnetization, by extending the Wohlfarth model for itinerant ferromagnets with volume-dependent terms [54]. Phenomenological Landau and Bean-Rodbell models, which are commonly used for a mean-field-type description of first-order magneto-structural transitions also presume $\omega \sim m^2$ [55]. Simultaneous measurements of the magnetization $m(H)$ and volume change $\omega(H)$ provide the opportunity to test this relation and evaluate the proportionality coefficient $kC_{mv}(T)$ (Fig 4(d)). Evidently, the experimentally obtained $\omega(m^2)$ is a non-monotonic function of the temperature and field. This implies that the magneto-volume effect is based on additional, more complex fundamental mechanisms.

A detailed picture of the incoherent changes occurring in the magnetic and structural subsystems of LaFe$_{11.8}$Si$_{1.2}$ under a field-induced, first-order phase transition is depicted in Fig 5(a,b). The solid lines in Fig 5(a) show the field dependencies of $m(H)$ (red) together with the volume changes $\omega(T)$ (blue), with all of them simultaneously measured in a magnetic field cycle 0.9→1.6→0.9 T and at a stable temperature in the cryostat $T_0$=186 K. In this experiment, the field-sweep rate was $10^{-5}$ T/s only, and the total measurement time was approximately 40 hours. The conditions in this experiment can be assumed to be near-isothermal, and the temperature change of the sample (bottom panel of Fig 5(a)) exhibits jumps at specific fields, where the field-induced heat release or the heat absorption occurs. The vertical dotted lines in Fig 5(a) indicate four selected fields: $h_1$ and $h_2$ for the two-stage PM-FM transition and $h_3$ and $h_4$ for the reverse FM-PM transition. We can see that at $h_1$ the PM-FM transition begins and the magnetization increases, but the volume of the sample reduces slightly, and this process is accompanied by heating of the sample. Between $h_1$ and $h_2$ the magnetization gradually increases, while the volume of the sample stays practically the same. Later, at $h_2$ the sample abruptly expands and a second sudden change in the magnetization occurs, accompanied by heating of the sample. During demagnetization, at $h_3$ the volume of the sample shrinks drastically, together with a 30% reduction in the magnetization and cooling of the sample. Between $h_3$ and $h_4$ the magnetization gradually decreases, and at $h_4$ the second stage of the transition completes, together with the second stage of heat abortion, and below $h_4$ the sample is in the paramagnetic state.

To stabilize the intermediate state where the volume of the sample corresponds to the paramagnetic state but the magnetization is around 1 $\mu_B$ per Fe-atom (for 186 K this field should be between $h_1$-$h_2$ or $h_3$-$h_4$, see Fig 5(a)), the envelope (major) and minor hysteresis loops were measured. In Fig 5(b) the scales of the $m$ and $\omega$ axes were adjusted to ensure that the plotted $m$ and $\Delta l/l$ values coincide in small and large magnetic fields above and below the transition. We can see that the major hysteresis loops (dotted lines in Fig 5(b), the field-sweeping rate was 5 mT/s) demonstrate no conjunction of $m(H)$ and $\omega(H)$ in the transition region, and the temperature change of the sample exhibits two-stage transitions. Solid lines show minor hysteresis loops 1.8 T→1.23 T→1.8 T measured with a 30-minute dwell before switching between demagnetization and magnetization. The minor hysteresis loops obtained in this way demonstrate that by interrupting the transition at $h_5$ where the volume starts to reduce (dotted line in Fig 5(b)) it is possible to stabilize the intermediate state, when the magnetization of the sample is about 50% of the magnetization in FM state, but the volume corresponds to its value in the paramagnetic state. This intermediate state is stable, and we experimentally confirmed no change in this intermediate state during a dwell of 24 hours.

We also carried out a series of similar experiments using various optimally annealed samples with different Si concentrations (LaFe$_{13-x}$Si$_x$ with x=1.2, 1.3, 1.4) and with distinct microstructures (specimens with grain sizes of 50–200 µm were prepared by induction melting, whereas samples with grain sizes of 10–20 µm were synthesized by arc melting with subsequent suction casting). Our simultaneous *macroscopic* measurements made on all these samples indicate the two-stage nature of PM-FM transition in LaFe$_{13-x}$Si$_x$ compounds with x<1.4.

We can see that simultaneous macroscopic measurements unambiguously confirm the two-stage nature of the transition. However, only macroscopic measurements are often insufficient to understand the true nature of the phenomenon and to reveal the key mechanisms governing the transition. For a deeper understanding of the essence of the multi-stage transformation, several element-selective techniques, such as x-ray absorption spectroscopy (XAS), x-ray magnetic circular dichroism (XMCD) and Mössbauer spectroscopy were applied in combination with parameter-free, first-principles calculations in the framework of the density functional theory (DFT). These enabled us to track the evolution of the electronic, structural and magnetic properties of $LaFe_{11.8}Si_{1.2}$ across the field-induced, first-order phase transition at an atomistic level.

Since simultaneous measurements of the macroscopic properties of $LaFe_{11.8}Si_{1.2}$ were carried out by increasing and decreasing the magnetic field, both magnetic and thermal hysteresis are clearly visible in Fig 3, 4(a,b,c) and 5(a,b). On the other hand, detailed XAS, XMCD and Mössbauer experiments were performed only in an increasing magnetic field, because the transformation of the electronic structure associated with the backward transition must occur in the reverse order, but fundamentally it has the same peculiarities. Thus the hysteresis phenomenon is not detected in Figs 5(c,d) and 6.

**C. XAS and XMCD measurements.**

XAS and XMCD spectra were measured at the Fe K-edge as a function of the applied magnetic field $\mu_0 H$ ranging from 1.65 T to 2.4 T. These spectra revealed significant changes in the electronic structure across the transition. As an example, the XAS (red line) and XMCD (black line) spectra measured at 187 K and in an applied field of 2.4 T are shown in the inset of Fig 5(c). To follow the changes in the XAS as a function of the applied field in more detail, the differences of the X-ray absorption coefficients diff(XAS)=$\mu(E,\mu_0 \cdot H)-\mu(E, 2.4\ T)$ are shown in Fig 5(c) for the different applied magnetic fields. These differences can be described by two effects: Firstly, the XAS intensity at photon energies below 7114 eV at small applied magnetic fields are lower as compared to the 2.4 T spectrum. Hence, the diff(XAS) signal in the photon energy regime of 7114 eV is largest (and negative) when inspecting the black line (difference signal $\mu(E,1.65\ T)-\mu(E, 2.4\ T)$) in Fig 5(c). This low-photon energy part of the XAS spectrum is typically assigned to quadrupole transitions of the 1s core electrons to the 3d states of Fe. Secondly, the absorption edge shifts towards lower photon energies with an increase of the applied magnetic field leading to a derivative like shape of the diff(XAS) signal. These two observations (intensity variation and energy shift of the XAS) point to more localized 3d states of the Fe above the phase transition, i.e. at larger magnetic fields. We have reproduced in Fig 5(d) the variation of the XAS intensity at photon energies of 7110.8 eV (red dots) and an energy shift of the X-ray absorption edge (blue dots) as a function of $\mu_0 H$. The dashed line in Fig 5(d) is a guide for the eye. This confirms that the described changes in the XAS spectra occur due to a two-stage transformation of the local electronic structure, while the transition observed in the XMCD signal (black line in Fig. 5(d)) is relatively smooth. In turn, the changes in the XAS spectra at photon energies above 7120 eV, where no transitions to the 3d states are expected and thus fully delocalized 4p states of Fe only are probed, do not show a two-stage behaviour. Therefore, XAS and XMCD, which probe the electronic structure and magnetic subsystem simultaneously, provide a first indication that the two-stage nature relates to the rather localized 3d states of Fe. Hence, the two-stage process originates on the atomic scale, rather than due to microstructural changes during the transition. To reveal the intrinsic nature of the mechanism, we need further measurements which can identify the evolution of magnetic moments and volume with atomic resolution.

### D. Mössbauer spectroscopy

Mössbauer spectroscopy differentiates the individual moments at the Fe-sites in terms of the $^{57}$Fe hyperfine fields $B_{hf}$[56]. In Fig 6(a) we plot the average distribution of $B_{hf}$ as a function of the macroscopic spontaneous magnetization $m$ measured in the temperature range 5–320K. The distribution $B_{hf}(m)$ reveals at least four distinct microscopic magnetic states with $B_{hf}$ =0 T, 10 T, 24 T and 30 T. In the disordered state and for small values of $m$, rapid fluctuations on the ps time scale between the different magnetic states and different orientations of the local magnetic moments lead to $B_{hf}$ =0 [48,49].

In the intermediate range 7 $\mu_B$/f.u <$m$< 22 $\mu_B$/f.u, fluctuations originating from the growth and reversal of mesoscopic magnetic clusters take place on a longer time scale. We observed the coexistence of two distinct maxima in the distribution around $B_{hf}$ ~10 T and ~24 T, which can be associated with two states having a magnetic moment $\mu_{Fe}$ proportional to $B_{hf}$. Approaching the saturation magnetization (i.e., at temperatures below 180 K) the average $B_{hf}$ increases steeply with $m$, finally reaching a value of 30 T in the FM phase. The significant stepwise changes in $B_{hf}$ are consistent with the large differences in the averaged effective moments $p_s$ and $p_c$ estimated from the fit to the Curie-Weiss law in the FM and PM phases, respectively.

### E. DFT calculations

The competition of two or more magnetic states of Fe in La(Fe$_{11-x}$Si$_x$)$_{13}$ alloys has been discussed before in terms of the theory of an IEM transition [34]. Previous DFT calculations [58,59] already confirmed that the energy surface $E(V,m)$ is characterized by metastable minima at different volumes $V$ and magnetizations $m$.

These differ in the local Fe-moments $\mu_{Fe}$ and are connected with a pronounced minimum electronic density of Fe *3d* states at the Fermi energy, which stabilizes the ground-state Fe-moment of about 2.2 $\mu_B$ and fills up as the local magnetic moment of Fe decreases with magnetic disorder [8,13,60]. Through adiabatic electron-phonon coupling this gives rise to the unexpected softening of the lattice in the PM phase due to substantial changes in interatomic force constants. These are fingerprinted in the Fe-partial vibrational density of states as obtained from nuclear resonant inelastic scattering (NRIXS) and DFT, both being in excellent agreement [8,13,20,61] demonstrating the predictive power of the theoretical approach.

After XAS and XMCD revealed the interplay between electronic structure and local magnetic moments in the two-step transition, DFT finally reveals the connection between the average magnetization *m* and site-resolved local moments at the Fe sites $\mu_{Fe}$ to the (relative) volume per atom. In Fig. 6(b) both the $\mu_{Fe}$ and *M* axes have quadratic scales, whereas the right-hand axis in Fig 6(b) is linear and corresponds to the spontaneous magnetostriction $\omega$. In this way, the proportionality of $\omega$ to $m^2$ and $\mu_{Fe}^2$ can be expressed with a straight line, between $m=20$ $\mu_B$/f.u ($\mu_{Fe}=1.8$ $\mu_B$), and the ground-state equilibrium magnetization $m=24.5$ $\mu_B$/f.u. *($\mu_{Fe}=2.2$ $\mu_B$)* is described by a comparatively steep slope (green triangles and dashed line in Fig. 6(b)). Below $m=20$ $\mu_B$/f.u. the slope decreases abruptly. This is related to the presence of magnetic disorder between essentially localized moments, which reduces *m* but not $\mu_{Fe}$. Still, we observed a slight decrease towards the PM state (*m=0*), which is due to the increasing admixture of Fe-atoms in low-spin states, which is also observed in the Mössbauer measurements in Fig. 6(a). These occur in particular on the Fe-I sites. Indeed, the experiment (Fig. 3(d)) shows a steep linear dependence of $\omega$ on $m^2$ at high fields as well, which disappears at low values of *H*. Thus, Fig. 6(b) tells us that a theory of moment-volume effects in La(Fe,Si)$_{13}$ should be based on the proportionality between $\omega$ and $\mu_{Fe}^2$ (instead of $m^2$), leading to the following scenario for the observed two-stage transition: below $M=20$ $\mu_B$/f.u., the changes in *m* originate from magnetic disorder (spin flips) [8,13]. Starting from the PM state, the applied magnetic field leads in the first line to a reorientation of the atomic moments with $\mu_{Fe}=1.8\mu_B$ and a decrease of the low-spin states. This causes a large increase in magnetization, in combination with a moderate change in volume. When nearly all the moments are FM ordered, we induce at some critical value $H_{cr}$ a discontinuous metamagnetic transition to the high-spin state with $\mu_{Fe}=2.2\mu_B$, which is accompanied by a large and discontinuous volume change. The proportionality between $\mu_{Fe}$ and *m* in Fig. 6(b) justifies the use of empirical Landau and Bean-Rodbell-type models with effective parameters chosen empirically [62,63] or derived from DFT [64], when modelling the phase transition in the range where magnetic order is not fully complete, but it also explains their failure to predict the second stage of the transition.

## IV. CONCLUSIONS

Thus, it is an intrinsic property of LaFe$_{11.8}$Si$_{1.2}$, i.e., the itinerant electron metamagnetism related to the Fe *3d* electrons, which enforces that the first-order transition does not complete in one but two distinct stages. Our study indicates reduced Fe-moments including the presence of quenched low-spin moments close to the paramagnetic state. Approaching the transition, the reduced magnetic moments are ordered at the first stage, which does not lead to a significant lattice expansion and is in experiment even associated with a slight contraction. Finally, the IEM transition to the high-spin state takes place (second stage), which increases the magnetization moderately, but is accompanied by large magneto-volume effect (Fig 7(a)).

Ultimately, our work shows a new pathway to disentangle the interplay between the structural, magnetic and electronic degrees of freedom, and serves as the next step towards a complete understanding the driving forces of the transition along with the origin of thermal hysteresis in magnetic phase-change materials (Fig. 7(b)). Since the transformations occurring in strongly interacting sub-systems of a solid can be very complex, and the roles and reciprocal contributions of these sub-systems to the total entropy change can be very specific, our simultaneous approach opens up new frontiers in this area. We suggest that the two-stage mechanism for the first-order transition, proposed here for La(FeSi)$_{13}$-type alloys, can be extended to a large class of advanced magnetic materials with itinerant moments exhibiting comparable field- and temperature-induced transformations, for example, Mn-Fe-P-Si, and Heusler-type alloys.

**Acknowledgments**

We acknowledge funding from the Deutsche Forschungsgemeinschaft (DFG, German Research Foundation), Project ID No. 405553726-TRR 270 HoMMage, Germany, from the German Federal Ministry of Education and Research (BMBF) under the grant number BMBF-Projekt05K2019 ULMAG and from European Research Council (ERC) under the European Union's Horizon 2020 research and innovation programme (Grant No. 743116—project "Cool Innov").

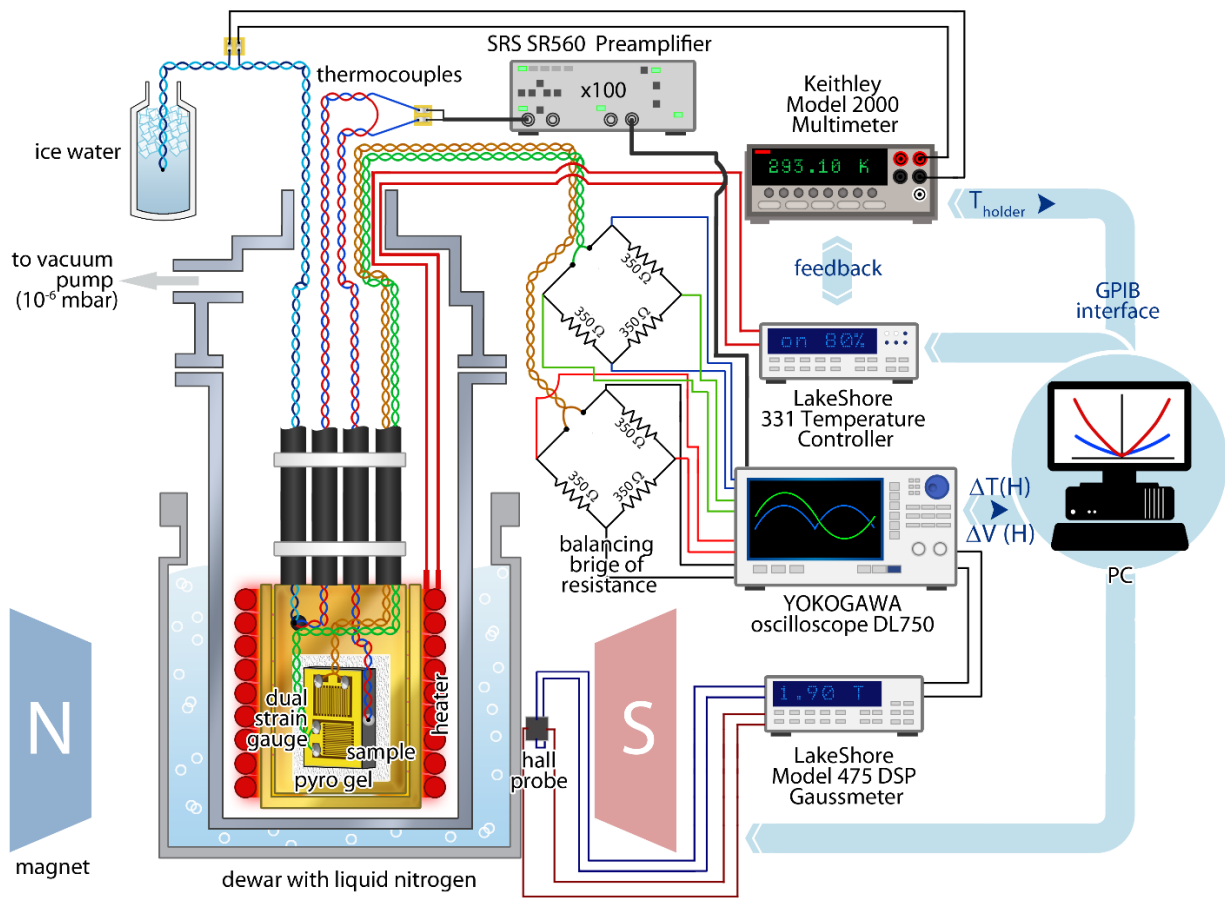

Fig 1. The purpose-built set-up for direct measurements of the adiabatic temperature change and both the longitudinal and the transversal adiabatic magnetostrictions.

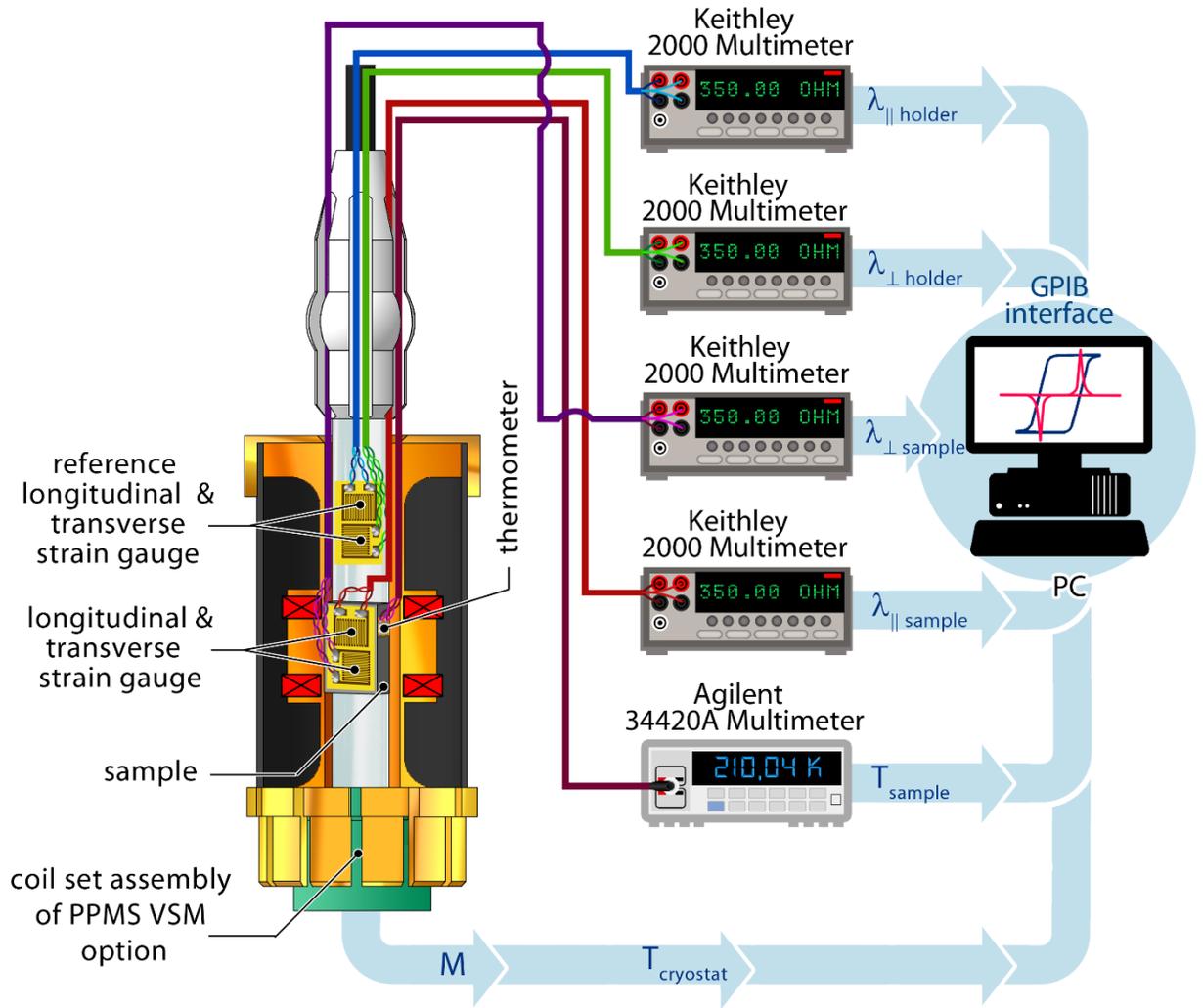

Fig 2. Purpose-built experimental setup for the simultaneous measurements of magnetization, magnetostriction and temperature change. This scientific instrument greatly enhances the measurement capacity of the commercial VSM option for the QD PPMS.

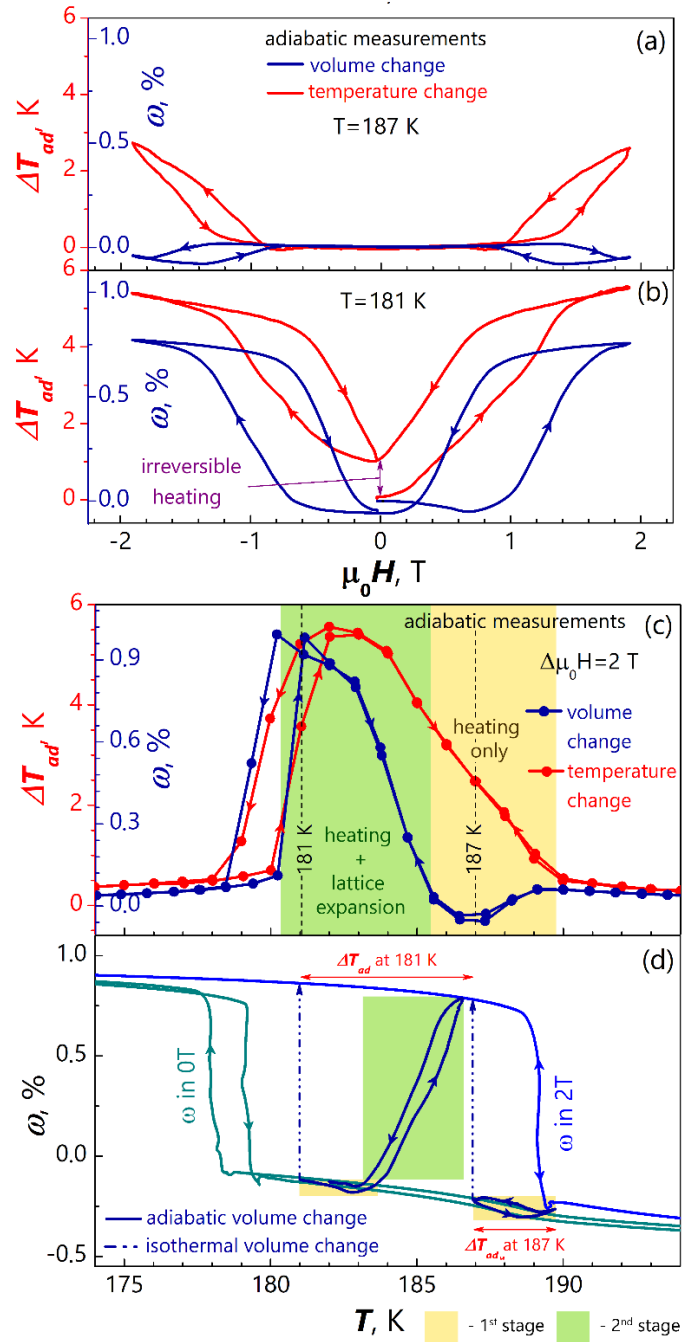

Fig 3. (a,b) Field dependencies of $\Delta T_{ad}(T)$ (red) and $\omega_{ad}(T)$ (blue) simultaneously measured at T=187 K (only temperature change) and at T=181 K (temperature change and lattice expansion); (c) Temperature dependencies of simultaneously measured $\Delta T_{ad}(T)$ (red) and $\omega_{ad}(T)$. The yellow shadowed area indicates the temperature range where the heating occurs without any lattice expansion, whereas the green area indicates temperatures where the lattice expansion and MCE occur simultaneously; (d) $\omega(T)$ diagram where adiabatic $\omega_{ad}(H)$ tracks are shown as blue lines and they demonstrate two-stage behaviour for starting temperature 181 K, but only temperature change for 187 K.

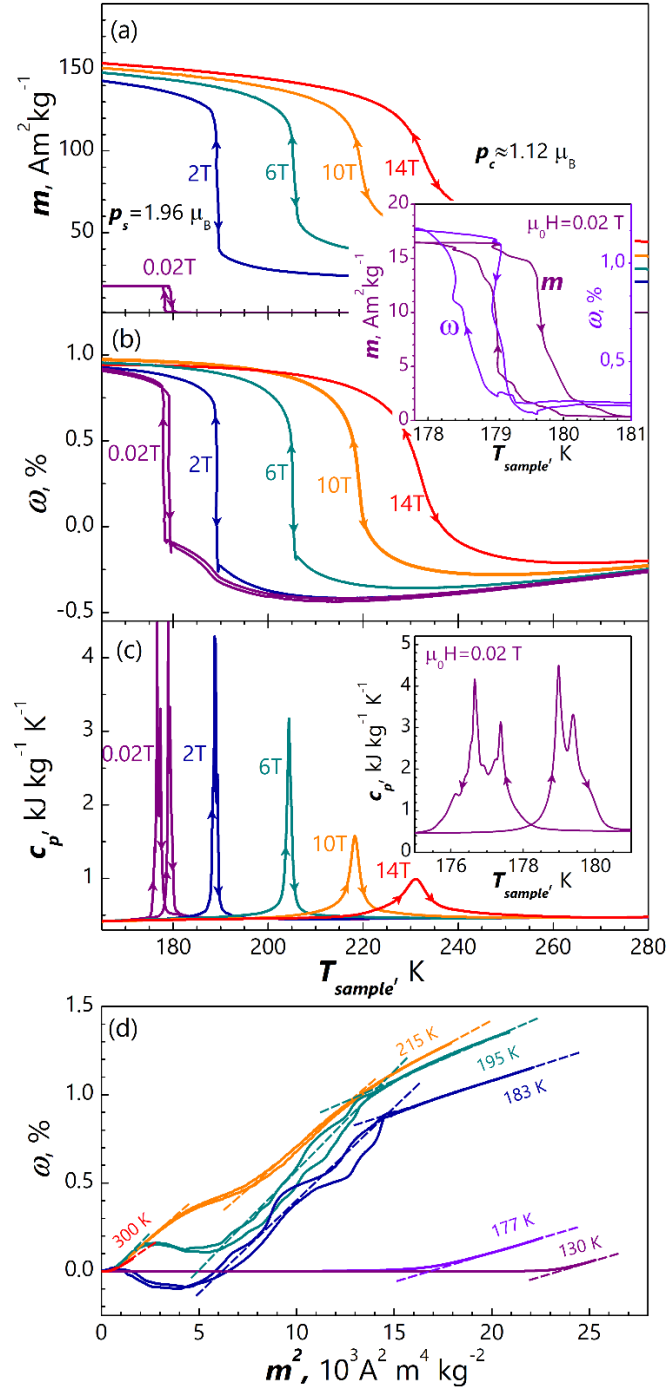

Fig 4. (a) Temperature dependencies of magnetization $m(T)$; (b) Temperature dependencies of volume change (inset shows $m(T)$, $\Delta\omega(T)$ measured in 0.02 T); (c) temperature dependencies of heat capacity $c_{p,H}(T)$, (inset shows $c_{p,H}(T)$ measured at zero field at the transition) and (d) the temperature dependencies of $\omega(m^2)$ obtained for different temperatures.

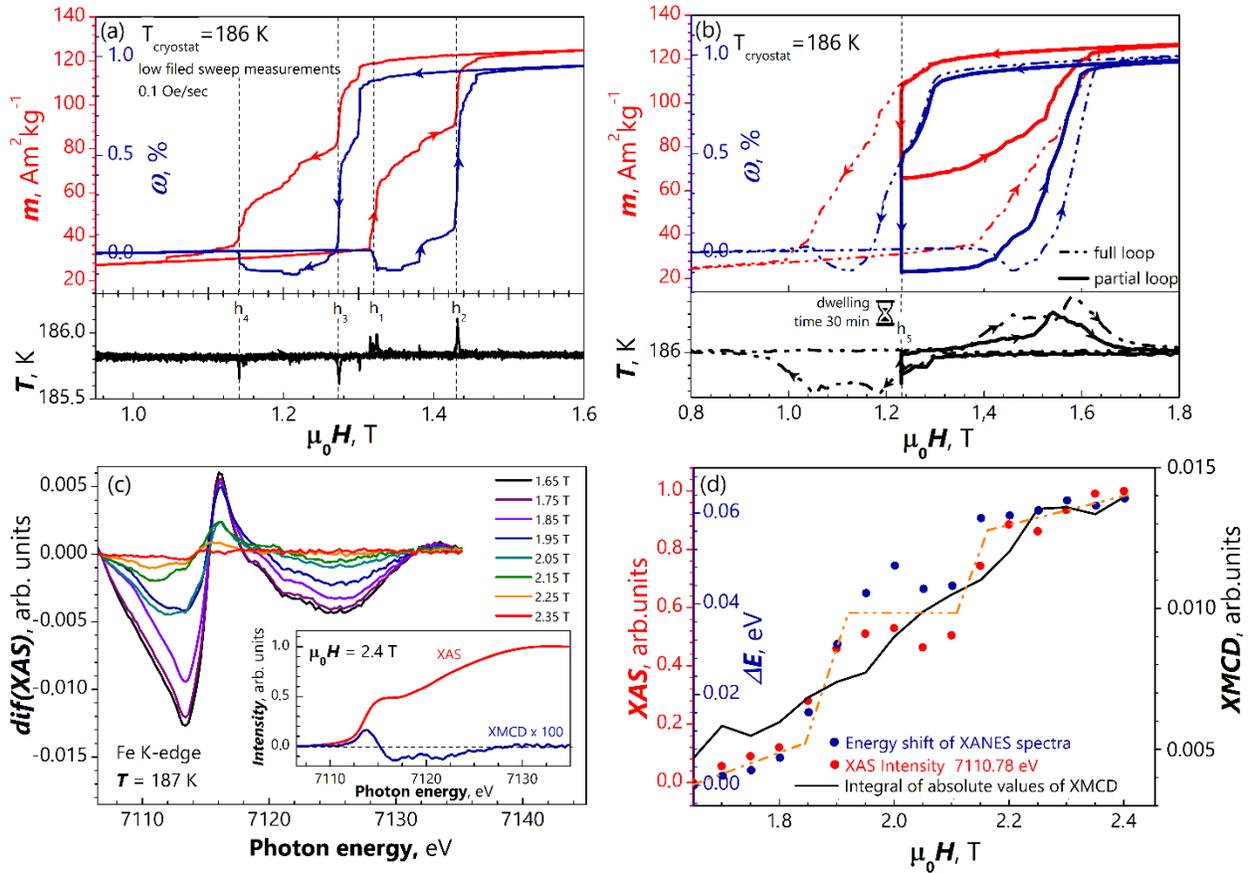

Fig 5. (a) *m(H)* (red) together with *ω(H)* (blue) quasistatically measured at *T*=186 K, the field sweep rate was 0.1 Oe/s. (b) Major (dashed line) and minor (solid lines) hysteresis loops for *m(H)* and *ω(H)*; minor loops were measured with 30-minute dwell between demagnetization and magnetization. At the bottom, one can see the temperature change of the sample. (c) Differences of the XAS spectra recorded at different field values with respect to the XAS spectrum recorded at 2.4 T (diff(XAS)=μ(E,μ$_0$·H)-μ(E, 2.4 T)). Inset shows the XAS (red line) and XMCD (blue line) spectra measured at 187 K and under an applied field of 2.4 T. (d) Variation of XAS intensity at photon energies of 7110.8 eV (red dots) and an energy shift of the absorption edge (blue dots) as a function of applied magnetic field. The solid black line depicts the magnetic field dependence of the integrated XMCD signal. The dashed line is a guide for the eye to follow the XAS signal together with the energy shift of the XAS data as a function of μ$_0$·H.

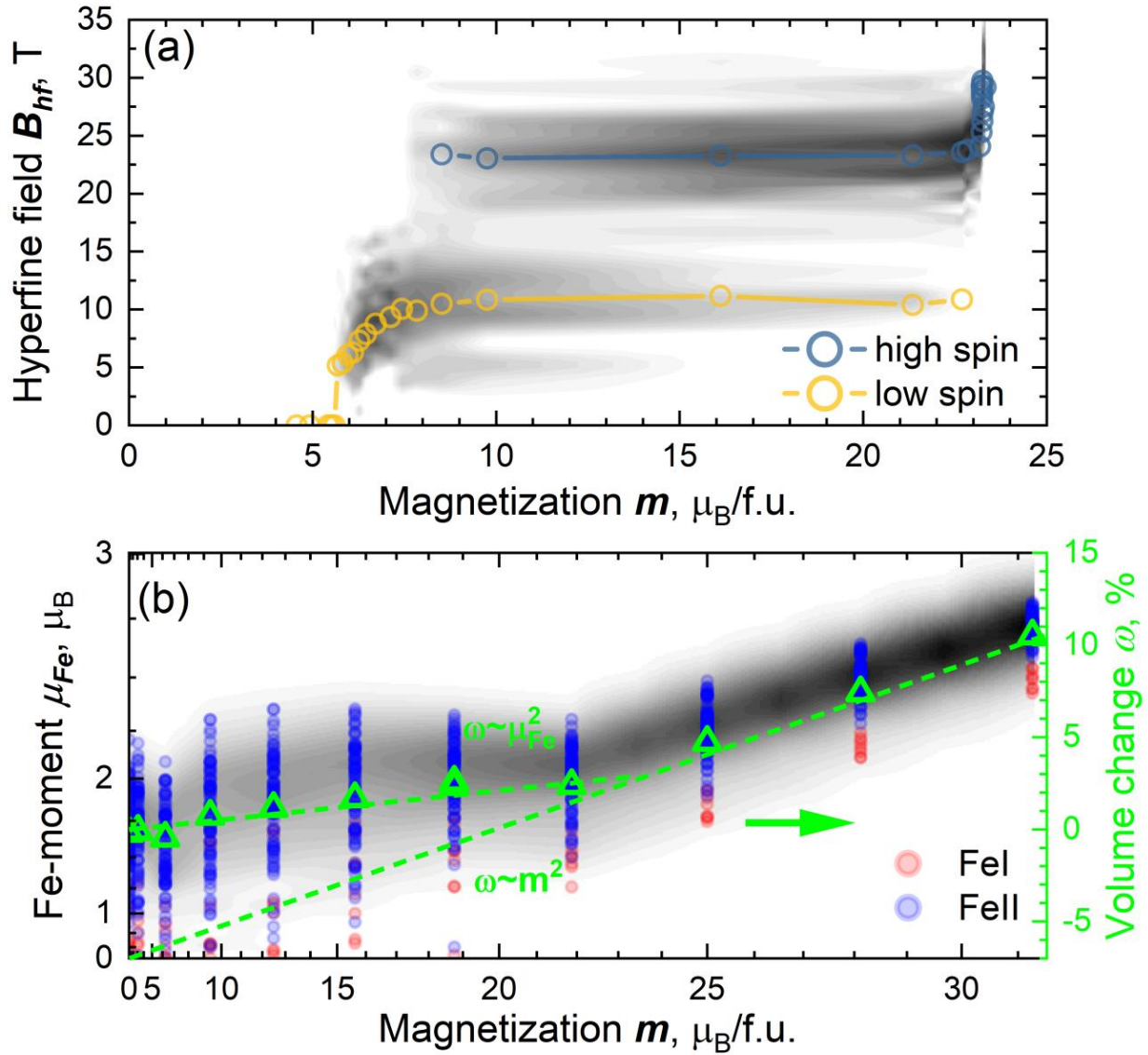

Fig 6. (a) Hyperfine field ($B_{hf}$) obtained from Mössbauer spectroscopy confirming the existence of high-spin (localized) and low-spin (itinerant) states. Icons symbolize the average $B_{hf}$ of the low-spin and high-spin contributions. (b) Absolute value of the local moment at the Fe sites $\mu_{Fe}$ vs. average magnetization $m$ per formula unit (f.u.) obtained from DFT calculations in a 112-atom cell. The red and blue dots indicate values of individual Fe moments at the Fe$_I$ and Fe$_{II}$ sites, respectively. The intensity of the underlying grey-shadow area visualizes the corresponding probability density as a guide to the eye. The right-hand axis refers to the relative volume change $\omega$ as a function of $m$ (green triangles). Both the moment and magnetization axes are scaled quadratically, such that the relations $\omega \sim \mu_{Fe}^2$ and $\omega \sim m^2$ appear as straight (dashed) lines.

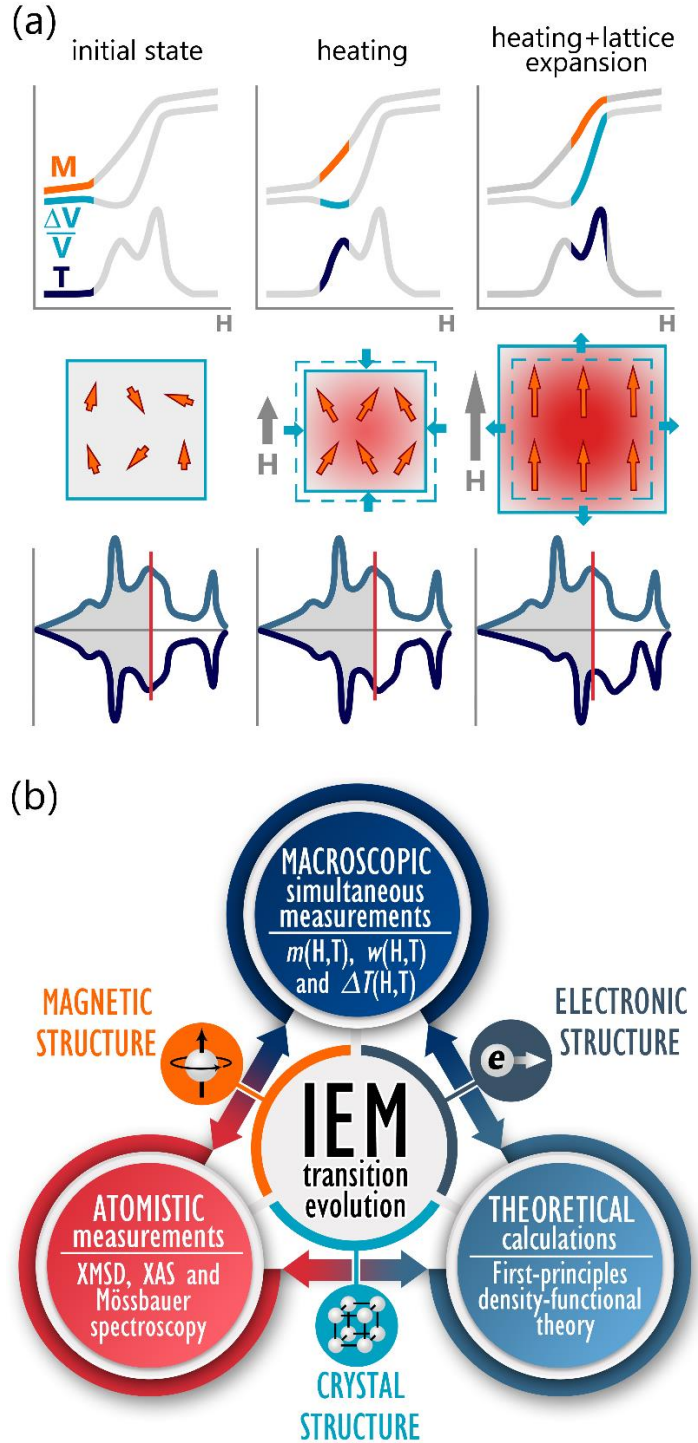

Fig 7. (a) Schematics of the two-stage transition in $LaFe_{11.8}Si_{1.2}$ alloy. (b) Simultaneous measurements of *macroscopic* parameters reveal the rich detail of the magneto-structural first-order transition. The experiments on an *atomistic* scale reveal the nature of the transition on a length scale below $10^{-8}$ m. All these experiments can be understood in terms *electronic structure* theory supplied by *density functional theory calculations* and can be used as training data for machine-learning algorithms.